\documentstyle[epsfig,mathptm,subfigure,amsmath,referee]{mn}

\title[Measuring the Hubble Constant]{Measuring the Hubble Constant
from Ryle Telescope and X-ray observations, with application to Abell
1413}

\author[K. Grainge et al]{Keith Grainge,$^{1}$ Michael E. Jones,$^{1}$
 Guy Pooley,$^{1}$ \cr Richard Saunders$^{1}$, Alastair Edge$^{2}$,
 William F. Grainger,$^{1}$ R\"udiger Kneissl$^{1}$ \\
 $^{1}$Astrophysics, Cavendish Laboratory, Madingley Road, Cambridge
 CB3 0HE\\ $^{2}$Department of Physics, South Road, Durham DH1 3LE}

\begin{document}

\maketitle

\begin{abstract}

We describe our methods for measuring the Hubble constant from Ryle
Telescope (RT) interferometric observations of the Sunyaev-Zel'dovich
(SZ) effect from a galaxy cluster and observation of the cluster X-ray
emission. We analyse the error budget in this method: as well as radio
and X-ray random errors, we consider the effects of clumping and
temperature differences in the cluster gas, of the kinetic SZ effect,
of bremsstrahlung emission at radio wavelengths, of the gravitational
lensing of background radio sources and of primary calibration
error. Using RT, {\sl ASCA} and {\sl ROSAT} observations of the Abell
1413, we find that random errors dominate over systematic ones, and
estimate $H_0 = 57^{+23}_{-16}$~$\rm km s^{-1} Mpc^{-1}$ for a an
$\Omega_M=1.0, \Omega_\Lambda =0.0$ cosmology.

\end{abstract} 


\section{Introduction}

Absolute distance measurements in astronomy are extremely valuable and
are very difficult to make. The extragalactic distance scale in
particular has profound implications for cosmology, but the
distance-ladder methods have generated long-standing
controversies. Direct methods of measuring cosmological distances, for
example via gravitational lensing effects, offer alternatives to the
distance ladder.  One such method exists through combining X-ray
observations of clusters of galaxies with the Sunyaev--Zel'dovich (SZ)
effect \cite{sandz}. This method takes advantage of the fact that the
SZ effect depends on a different combination of the intrinsic
properties of the cluster gas (density, temperature, physical size)
than the observed X-ray flux: hence the physical size of the cluster,
and via the angular size, its distance, can be measured. To see how
the distance depends on the measurable quantities, consider a blob of
gas of size $l$, electron number density $n_e$, at temperature
$T_e$. The observed X-ray flux density is
\[ F_X \propto n_e^2 f(T_e) l^3D_l^{-2}\label{Xbrem}\] where $D_l$ is
the luminosity distance 
and $f(T_e)$ is the temperature dependence of the X-ray emissivity (including
Gaunt factor). The SZ effect 
from the same gas is
\[ \Delta T_{\rm SZ} \propto  n_e T_e l. \label{SZeffect}\]
If the blob subtends an angle $\theta$ then
\[ \theta = l/D_a,\]
where the angular size distance $D_a = D_l (1+z)^{-2}$. Rearranging, 
\[ H_0 \propto D_a^{-1} \propto \frac{T_e^2 F_X}{\Delta T^2 f(T_e)
\theta}, \label{calcH0}
\]
where we have dropped the explicit $z$ dependence. $H_0$ can therefore be
determined for a cluster of known redshift provided we can measure the gas
temperature, the X-ray flux and the SZ effect, and provided we can model the
distribution of gas in the cluster.

This method (e.g. Cavaliere et al \shortcite{cavaliere}) was suggested
shortly after the SZ itself was proposed, and since the first
successful detections of the SZ effect in the mid-1980s several
authors have published determinations of $H_0$ based on it, using both
single-dish (e.g. \cite{birk94,myers}) and interferometric (e.g.
\cite{jones-capri,reese}) SZ measurements. Despite the advantages of
the SZ method in by-passing the entire distance ladder, it does have
its own sources of error, which must be carefully addressed. These
include:

\begin{enumerate}
\item{ Beam switching that does not go fully outside the cluster;
contamination by radio sources in the cluster or (for single-dish SZ
measurements) in the reference beams; resolving out most of the total
flux of the cluster (for interferometric measurements); poor spatial
sampling of the cluster.}
\item{Errors in the temperature measurement of the cluster, due to
poor photon statistics, temperature substructure and clumping of the
cluster gas}
\item{Errors due to fitting a simplified smooth model of the gas
distribution to a real complicated cluster}
\item{Biases due to selection effects in the chosen cluster sample.}
\end{enumerate}

In this paper we describe our analysis techniques for determining
$H_0$ by combining SZ observations from the Ryle Telescope~(RT) with
X-ray data from {\sc ROSAT} and {\sc ASCA}. We apply this to data for
the cluster A1413, for which we have previously published SZ results
\cite{grainge96}. Finally, we consider the various sources of error,
both random and systematic, which affect our estimate.

\section{Cluster modelling} 
\label{model}

We use our data to measure the Hubble constant by creating a numerical
model of the temperature and density distributions in a cluster
deduced from X-ray data, and then using these to generate mock RT data
that can then be compared with real observations. The gas density is
modelled as a King profile in one or three dimensions (depending on
whether the cluster shows significant asphericity), and arbitrary
temperature profiles can be specified. The cosmological parameters
$H_0$, $\Omega_M$ and $\Omega_\Lambda$, upon which the relative scaling of the
X-ray and SZ data depend, can be adjusted until both sets of mock data
agree with the observations. The method is essentially similar to
those used in other SZ measurements of $H_0$, especially other
interferometric measurements \cite{reese}. Unlike Reese et al we do
not use the SZ data to constrain the shape parameters, as the X-ray
image contains far more information and would dominate any joint fit;
we do however allow for a non-spherical cluster.

\subsection{Physical model}

Our modelling is based on the analytical King
approximation~\cite{CF76} for the potential due to a self-gravitating
isothermal sphere, which gives results consistent with X-ray images,
at least in the inner parts of the gas distribution. The cluster total
density $\rho$ at radius $r$ from the cluster centre is given by
\[
\rho(r)=\rho_0 \left(1+\left({r/r_c}\right)^2 \right)^{-3/2} ,
\]
where $r_c$is the core radius of the cluster. Assuming that the gas
density,~$\rho_g$ is related to the total mass density by $\rho_g \propto
\rho^{\beta}$, then
\begin{equation}
\rho_g(r)=\rho_{g0} \left(1+\left({r/r_c}\right)^2 \right)^{-3\beta/2}
.
\label{kingden}
\end{equation}
The average value of $\beta$ determined by fits to X-ray surface brightness of
a large number of clusters is found to be $<\beta_{fit}>=0.67$ \cite{JF84},
but we fit for $\beta$ separately for each cluster.

Although we typically use isothermal models for fitting X-ray and SZ
data, we can vary the temperature to investigate specific effects.
We do not vary the X-ray emissivity constant in these
models; however, for typical rich cluster temperatures (7--10 keV) this
constant is a weak function of temperature when observing in the {\sl
ROSAT} 0.5--2~keV band.

\subsection{Simulating observations} \label{simmodel}

We generate two-dimensional images, of typically $256 \times 256$
pixels, of the model skies in SZ and in X-rays. This results in a
spatial resolution in the model of $\sim 50$~kpc and a total
simulation size of $\sim 10$~Mpc -- numerical experiments show the
results to be insensitive to these parameters. We can use either a
spherical King model (equation \ref{kingden}) for the density, or a
3-d modification of it in which core radii ($r_{cx}, r_{cy}, r_{cz}$)
are specified in three perpendicular directions, the directions
relative to the plane of the sky being specified by three Euler
angles. The density at a position ($x, y, z$) in these coordinates is
thus given by \[ \rho =
\rho_0(1+x^2/r_{cx}^2+y^2/r_{cy}^2+z^2/r_{cz}^2)^{-3\beta/2}\] We
integrate the functions shown below along the rows of a
three-dimensional array to generate the SZ brightness temperature and
X-ray images: 
\begin{equation}
\Delta T(i,j) = \sum_k n(i,j,k) T(i,j,k){{g(x)}\over{x^2}} {{k_B T_0
\sigma_T}\over{m_e c^2}} L ,
\label{szeffect}
\end{equation} 
\begin{equation*}
g(x) = {{x^4 e^x}\over{(e^x-1)^2}} \left( x \coth{x\over2}-4 \right),
\end{equation*}
\begin{equation*}
x = {{h \nu}\over{k_B T_0}} ,
\end{equation*}
\begin{equation}
{\rm X}(i,j) = \sum_k K n^2(i,j,k) t_{exp} L^3 , 
\end{equation} 
where $T_0$ is the CMB temperature, $t_{exp}$ is the exposure time of
the X-ray image and $K$ is a telescope, redshift and
cluster-temperature dependent constant determining the received count
rate, incorporating the bremsstrahlung and line emissivities, hydrogen
absorption, detector bandpass and K-correction, expressed as {\sl
ROSAT} PSPC counts per second due to a volume of $1 \rm m^3$ of gas,
of electron density $1 \rm m^{-3}$, at a luminosity distance of $1$
Mpc. $L$ is the size in metres corresponding to the angular extent of
each pixel $\theta_{cell}$; this is calculated using the cluster
redshift, $z$, and assumed values of $H_0$,  $\Omega_M$ and
$\Omega_\Lambda$.   

We have checked that we obtain results consistent
with the analytic expressions for both the X-ray and SZ profiles for
the spherical isothermal King model.  We choose not to use the
analytic expressions in our code since the approach described above
can more easily be modified to allow different distributions of gas
density and temperature. For example, we have been able to investigate
the effect of non-isothermal temperature profiles, and also central
cooling flow regions. The latter have no effect on the SZ signal,
however, since cooling flows are considered to be in pressure
equilibrium with the rest of the cluster.

\begin{figure}
\begin{center}
\leavevmode
\hbox{%
\epsfig{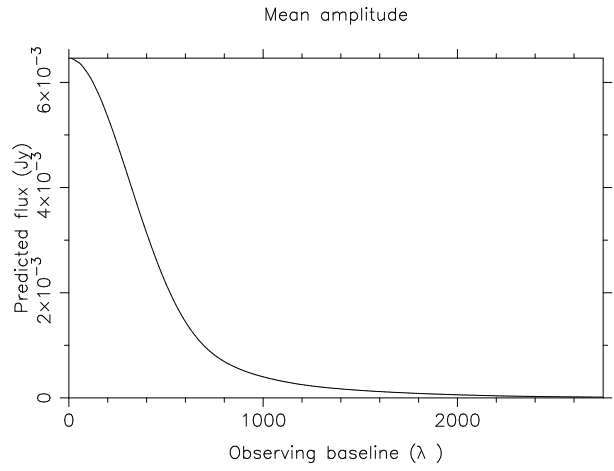}}
\end{center}
\caption{Simulated SZ flux as a function of projected RT baseline.
The cluster model used for this simulation was the best fit to the
{\sl ROSAT} map of A1413 (see section~\ref{a1413xfit}). During the course of a
12~hour run the RT observes with projected baselines $b>640\lambda$;
baselines shorter than this are shadowed by other antennas and the
data are flagged.}
\label{pmax}
\end{figure}

The brightness temperature decrement map is converted to flux density,
multiplied by the RT primary beam response, and Fast Fourier
Transformed to obtain the simulated aperture-plane response of the RT to
the SZ decrement. The best-fit cluster parameters found for the
cluster A1413 (see section~\ref{a1413xfit})
give a profile in the aperture plane for different
observing baselines shown in Figure~\ref{pmax}. Mock data that
correspond to the real RT observations can then be generated by
sampling this array at the observed coordinates in the aperture plane.

We also simulate the process of observing the sky with the {\sl ROSAT}
PSPC or HRI.  The model X-ray sky is convolved with the X-ray
telescope point-spread function, and a suitable background level
added. For making model fits a smooth background is used; we
can also make images in which each pixel value (signal plus
background) is replaced with a sample from a Poisson distribution of
the same mean, thus generating realistic X-ray images.

We tested the program to ensure that it gave the expected dependence upon
central number density, $\beta$, core radius, gas temperature and value for
$H_0$ and to check that the results for the simulations were not dependent
upon pixel size or overall array size.

\subsection{Predictions of SZ observations}\label{predict}

We have predicted the effect that changes in various cluster
parameters would have on our RT observations. In particular we were
concerned about the behaviour of the flux density measured on the
shortest RT baselines, since these are the most sensitive to the SZ
effect. (The shortest physical RT baseline is $870~\lambda$, but at for
lower declinations this can be projected down to a minimum of the
antenna diameter of $640~\lambda$.) 
The results here were obtained
using the model parameters for Abell~1413, but are
typical of all the rich clusters we observe.

\begin{figure}
\begin{center}
\leavevmode
\epsfig{angle=270,width=80mm,file=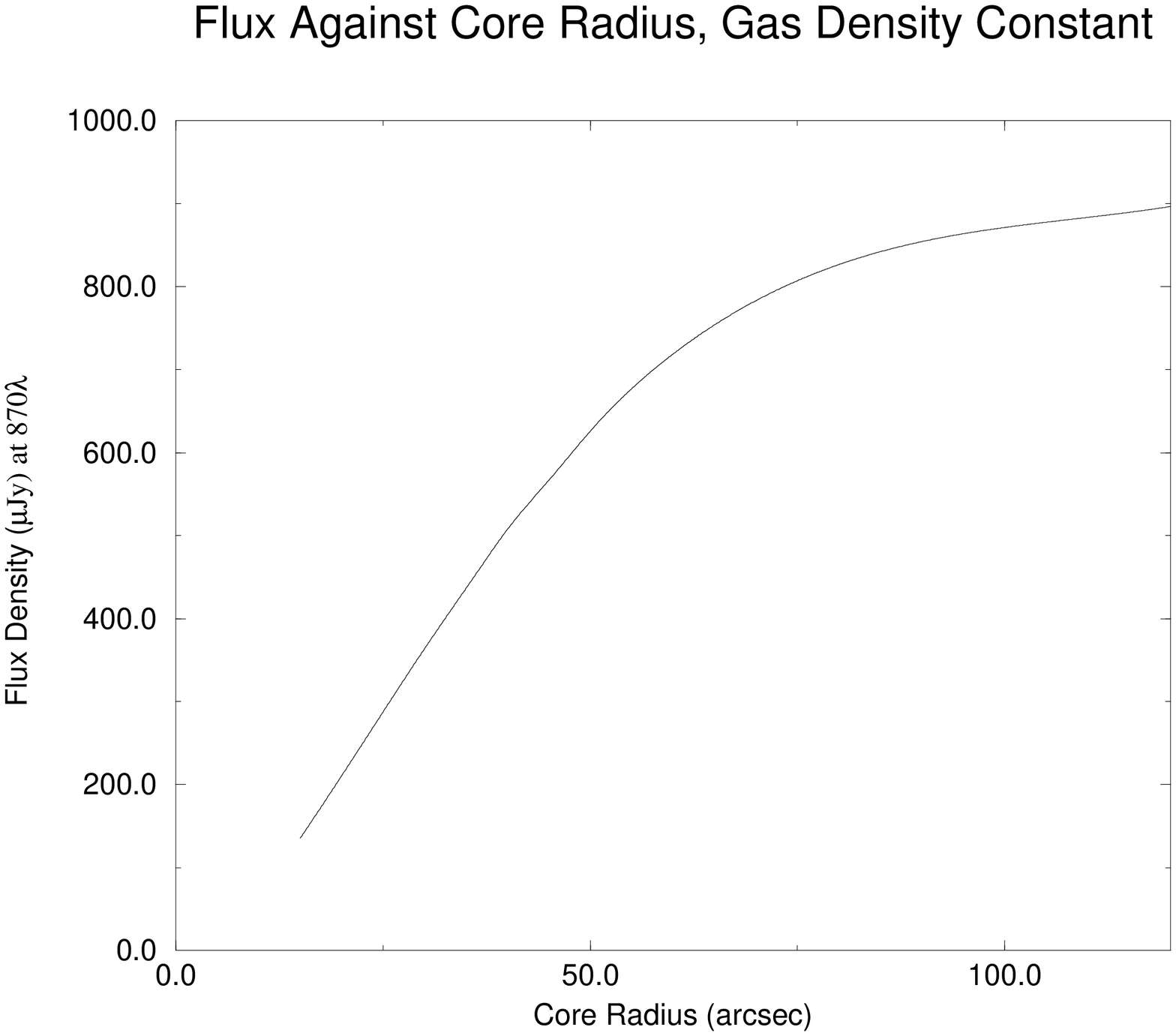}
\epsfig{angle=270,width=80mm,file=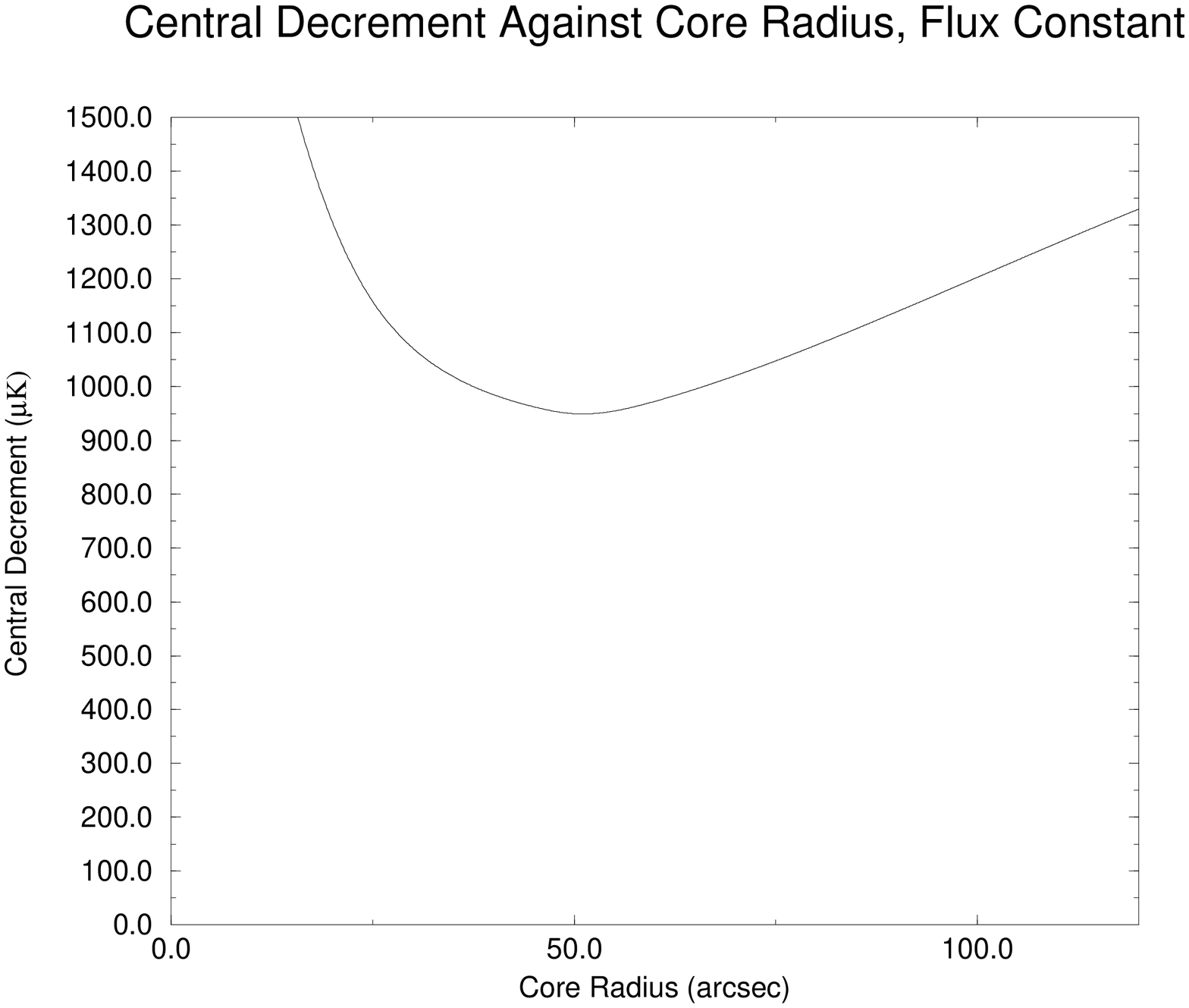}
\end{center}
\caption{(Left) Observed flux density on a baseline of 870~$\lambda$
against cluster core radius, with 
constant central gas density and temperature. (Right) Central
decrement against core radius, with constant observed flux density of
550~$\mu$Jy on a baseline of 870~$\lambda$, showing the relative
insensitivity of observed flux density to central decrement at this
resolution.}
\label{flxvrad}
\end{figure}

We simulated a cluster with using the best-fit parameters that we have found
for A1413 (see section~\ref{a1413xfit}) but varied the core radius. 
We found that the
central decrement is directly proportional to the core radius, as
expected, and that the flux density measured on the 870~$\lambda$
baseline quickly began to flatten off (see Figure \ref{flxvrad}) as
the RT starts to resolve out the SZ signal. This resolving out starts
to become significant for core radii of greater than about
70~arcseconds for the 870~$\lambda$ baselines.

In a further test
 we varied the central number density, $n_0$, for
various core radii to give the same observed flux density using a $870
\lambda$ baseline, and recorded the central temperature decrement that
these models would give. The results are also shown in
Figure~\ref{flxvrad}. The central decrement does not vary greatly
with core radius, especially around 60~arcseconds which is the
expected size for a cluster with core radius of about 200~kpc at moderate
redshift.  

\begin{figure}
\centering
\mbox{
\subfigure[Projected baseline = 870 $\lambda$]
{\epsfig{figure=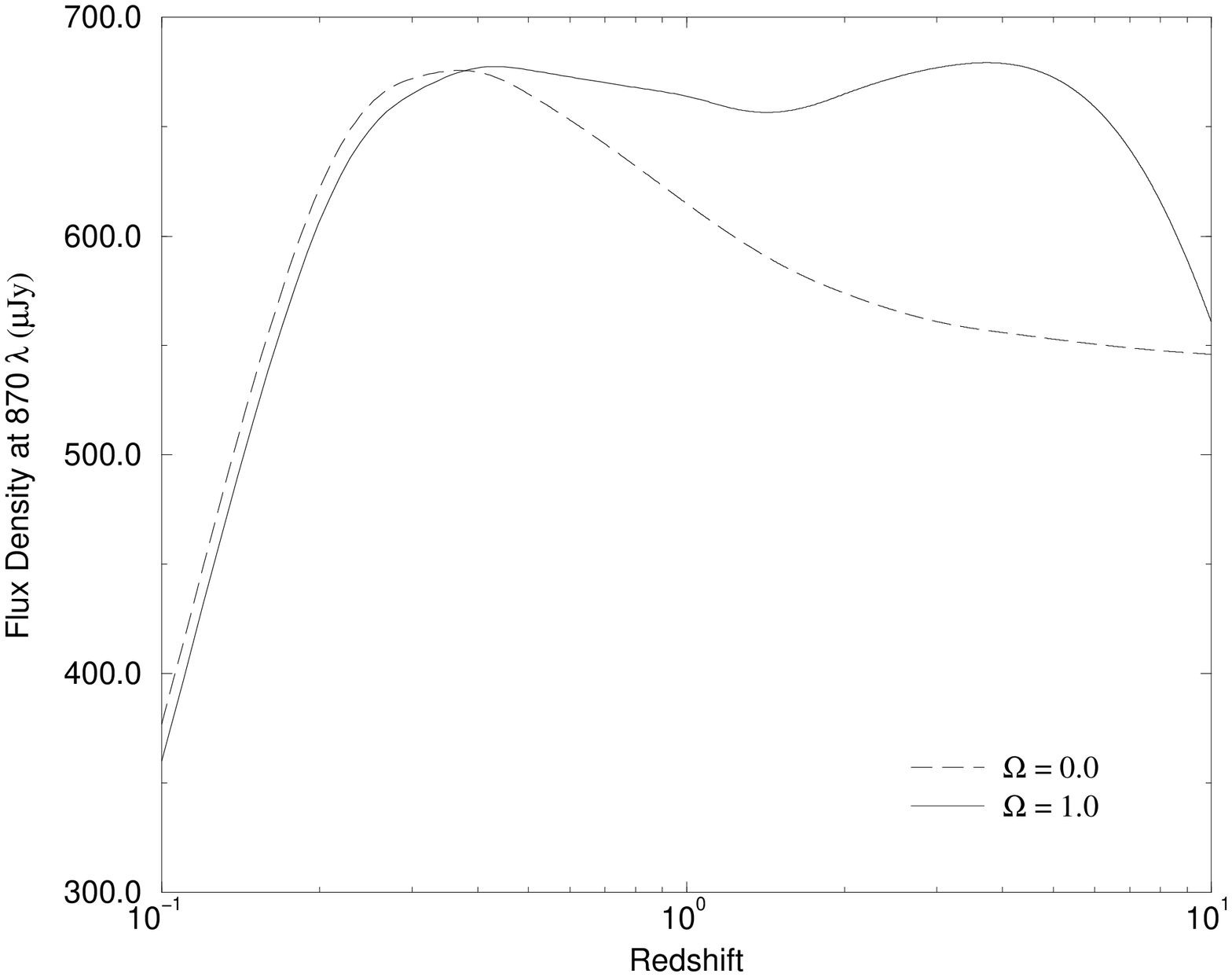,angle=270,width=80mm}}
      \subfigure[Projected baseline = 1740 $\lambda$]
{\epsfig{figure=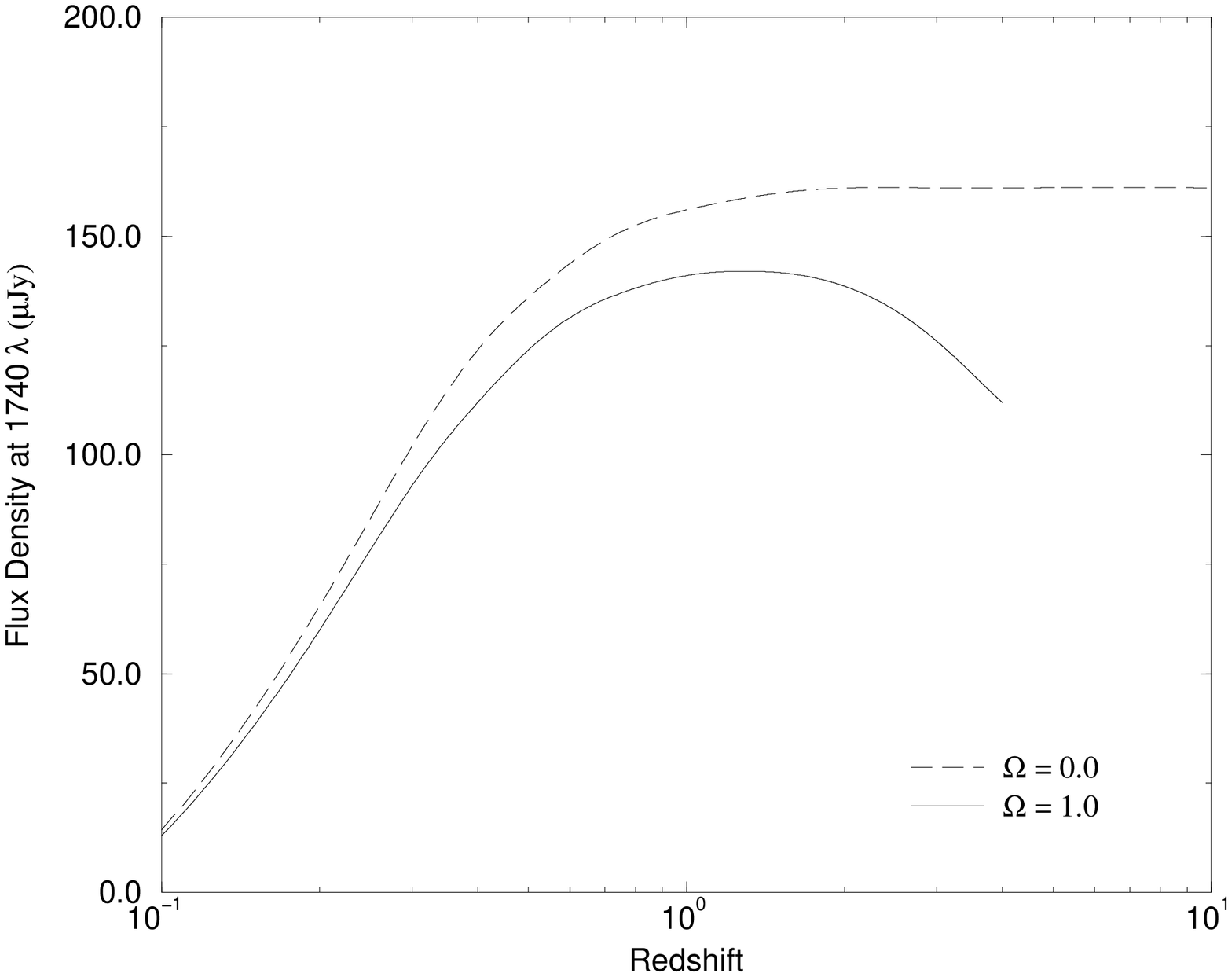,angle=270,width=80mm}}}
\caption{Flux observed for the cluster with core radius of 250~kpc
projected back in redshift,
in Einstein-de Sitter and Milne universes.}
\label{flx900}
\end{figure}

We have also assessed the effects of redshift and cosmology on the SZ
signal seen.  We simulated observations of clusters with the same
intrinsic parameters lying at different redshifts.  First we
calculated the clusters' angular core radii, corresponding to a
constant physical radius of 250~kpc, assuming Einstein-de Sitter
($\Omega = 1.0$) and Milne ($\Omega = 0.0$) cosmologies. The results
for observations with baselines of 870 and 
1740~$\lambda$ are shown in Figure~\ref{flx900}. 
It can be seen that
the RT has similar sensitivity to a cluster at redshifts between 0.2
and 10 on the shortest baseline in both cosmologies.
On the 1740~$\lambda$ baseline the signal
peaks for clusters at redshifts between 0.3 and 4 in an Einstein-de Sitter
cosmology, but tends to a maximum value in a Milne universe beyond a
redshift of 1.
We therefore
conclude that observations of the SZ effect with the RT should be
possible for clusters at any redshift from $z=0.1$ to $z=10$.

\begin{figure}
\begin{center}
\leavevmode
\hbox{%
\epsfig{angle=270,width=80mm,file=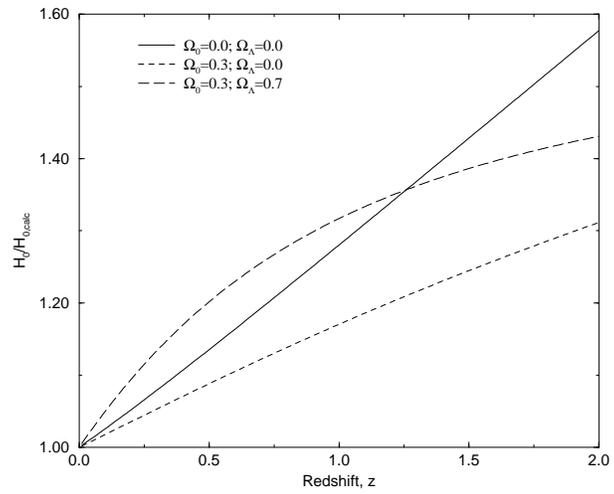}}
\end{center}
\caption{Correction factor to calculated value of $H_0$ for different
cosmologies, relative to an Einstein-de Sitter model.}
\label{q0corr}
\end{figure}

Although the SZ effect predicted to be detected by the RT from a
cluster of given physical size is largely independent of cosmology,
the value of $H_0$ that we will calculate from observations will be
affected by the values of $\Omega_M$ and $\Omega_\Lambda$ that we
adopt. The true value of $H_0$ 
and the calculated value $H_0^{\rm calc}$ will be related by
\[ H_0 = {{H_0^{\rm calc} D_a^{\rm calc}}\over{D_{a}}} \] where $D_a^{ \rm
calc}$ and $D_{a}$ are the calculated and true angular distances to the
cluster respectively. The correction factor $H_0^{\rm calc} / H_0$ is
shown in Figure~\ref{q0corr}.

\subsection{Fitting to X-ray images}
\label{x-fitting}

Since {\sl ROSAT} X-ray images contain far more information about the
gas distribution in a cluster than RT SZ data, we use the X-ray image
to model the density distribution. We assume the cluster is a triaxial
ellipsoid with two principal axes in the plane of the sky, and with
the core radius along the line of sight equal to the geometric mean of
the other two core radii. We then use a downhill simplex fitting
method, implemented by the routine {\sc amoeba}~\cite{numrep}, to
optimise the central density, core radii, ellipsoid centre, position
angle and $\beta$ parameter. The likelihoods of the data given the
model are calculated as the product of the Poisson likelihoods of
obtaining the observed count at each pixel in the X-ray image, given
the parameters. It is necessary to use the Poisson statistic rather
than $\chi^2$ in order to avoid biases in the regions of low count
rate where the two statistics differ significantly. This method also
allows arbitrary regions to be excluded from the fit; this is used for
example to exclude point sources in the X-ray image.

We use vignetting-corrected X-ray images (i.e. divided by the image of
effective exposure time) for the fitting. Although this results in a
distortion of the photon statistics across the map (partly due to the
presence of the non-vignetted instrumental background), this effect is
small in the central part of the image where all the cluster signal
lies. We also use images that are binned so as to under-sample the PSF
(pixel size equal to FWHM of PSF) so that the position dependence of
the PSF is negligible over the region of interest.

\begin{figure}
\epsfig{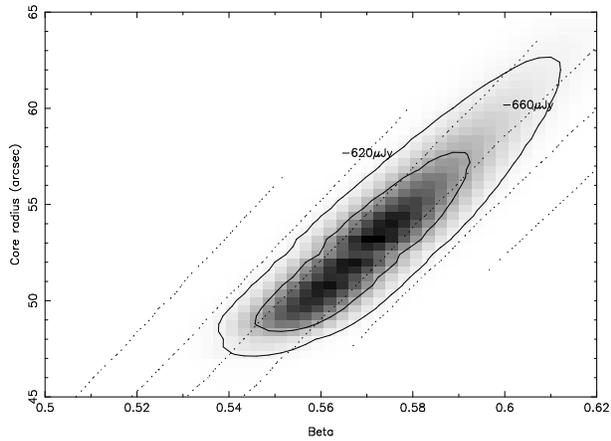}
\caption{Greyscale and solid contours: likelihood of fit to X-ray
data, as a function of $\beta$ and core radius of the major axis,
marginalised over $n_0$ but with the axial ratio and position angle
(which are very well determined) fixed. Contours are 67\% and 95\%
confidence. Dotted contours: predicted mean flux density on the
shortest RT baseline, for $h=0.5$. The degeneracy between $\beta$ and
the core radius clearly has little effect on the predicted flux
density, and hence on the fitted value of $H_0$. }
\label{degenerate}
\end{figure}

There is a degeneracy between $\beta$ and $\theta_c$, and $n_0$, in
that almost equally good fits can be obtained by changing the value of
$\beta$, and re-fitting for $\theta_c$ and $n_0$. However, although
the resulting value of the central temperature decrement is changed
for these different fits, the flux density observed by the
interferometer is almost unchanged. This is because the observed flux
density depends on both the central decrement and the shape of the
cluster, and on our most sensitive baselines the two effects almost
cancel. The effect of this can be seen in Figure~\ref{degenerate},
which shows contours of likelihood of the X-ray fit in the
$\beta$--$\theta_c$ plane overlaid with contours of the predicted flux
density observed by the RT. The flux density contours (which are used
to calculate $H_0$) lie parallel to the $\beta$--$\theta_c$ degeneracy,
which thus has little effect on the value of $H_0$ found. Reese et
al.~\shortcite{reese} find a similar result for the OVRO/BIMA arrays.

\subsection{Fitting to SZ data}

Interferometers measure the Fourier transform of the sky brightness
distribution at discrete points defined by the orientation of the
baselines, with independent noise values at each point. Inverting the
data to form an image results in both signal and noise being convolved
with the Fourier transform of the sampling function (the synthesised
beam), resulting in long-range correlations across the map. With
well-filled apertures and high signal-to-noise, this is not a
problem. Since the RT has few baselines on which the SZ effect is
detectable, these correlations are significant, ie the synthesised
beam has large far-out sidelobes. It is therefore very difficult to do
meaningful fitting to the data in the image plane. Accordingly, we fit
our data to models in the aperture plane, although we also make maps,
particularly in order to identify positive sources in the field.

Once a fit has been made to the X-ray image using an assumed value of
$H_0$, an SZ sky model and simulated RT aperture are generated as
described above. Variations in $H_0$ result in a simple scaling of the
amplitude of the SZ signal. The calibrated, source-subtracted RT
visibility data are read in and a $\chi^2$ statistic calculated
between the real and mock data as a function of amplitude scaling;
we  multiply by a prior that is uniform in log space (since $H_0$ is
a scale parameter) to find the peak and extent of the posterior
probability distribution.

\section{Calculating $H_0$ from A1413} \label{calchub1413}

A1413, at a redshift $z=0.143$, is a massive and highly luminous
cluster with a {\sl ROSAT}-band luminosity of $2.0 \times 10^{38}$ W
\cite{allen95}. We have previously published SZ data on the cluster
Abell~1413 \cite{grainge96} but did not derive an $H_0$ value there
due to the significant ellipticity of the cluster and the presence of a
moderate cooling flow, both of which complicate the process. Here we
calculate $H_0$ from A1413 taking both these effects into
consideration. As well as our RT SZ data, we use an observation made
with the {\sl ROSAT} PSPC, and a temperature derived from {\sl ROSAT}
HRI and {\sl ASCA} data.

\subsection{X-ray image fitting}
\label{a1413xfit}

\begin{figure}
\begin{center}
\leavevmode
\hbox{%
\epsfig{width=80mm,file=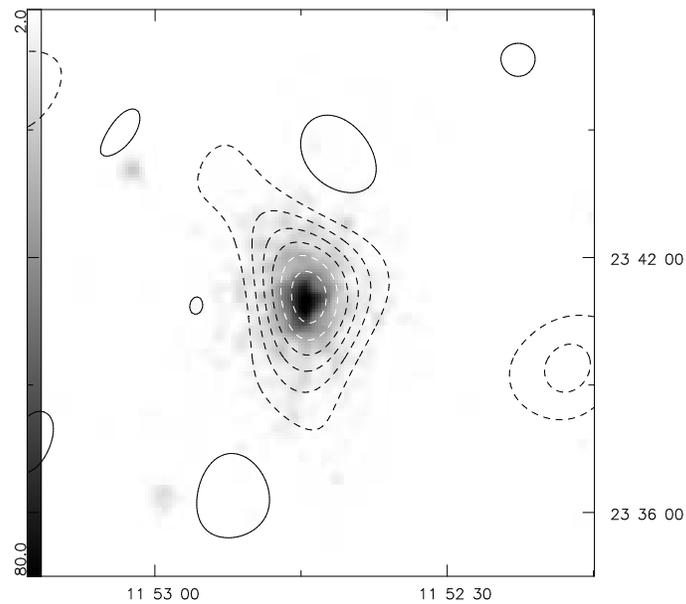,angle=270,clip=}}
\end{center}
\caption{X-ray and SZ images of A1413, from Grainge et al (1996). The greyscale is the {\it {\sl ROSAT} PSPC} image; it
runs from 2 to 80 counts in a $15''\times 15''$ pixel.  The contours show the
{\sc Clean}ed map of the naturally weighted RT 0--1~k$\lambda$ visibilities
with restoring beam $169''\ \times \ 110 ''$ at position angle 2.7$^{\circ}$;
contour levels at $-450$, $-375$, $-300$, $-225$, $-150$, $-75$, 75 $\mu$Jy.\label{A1413-0-1}}
\label{a1413-0-1}
\end{figure}

We fitted an ellipsoidal model (as described in section
\ref{x-fitting}) to the {\sl ROSAT} PSPC image observed on 1991
November 27, which has an effective exposure of 7696~s. We use only
the hard (0.5--2~keV) data in order to minimise the effect of Galactic
absorption; the image is shown in Figure~\ref{A1413-0-1}, along with
the RT SZ image. Allen et al.~\shortcite{allen95} find that A1413 has a
cooling flow with a mass deposition rate of 200 M$_{\odot} \rm
yr^{-1}$ within a radius of 200 kpc of the centre, and that the
temperature in this region is depressed by some 20\% from the value in
the outer regions. Therefore we excluded the X-ray data within the
cooling flow radius from our analysis, along with several X-ray point
sources which appear in the image. Keeping the cooling flow region in
the fit would result in a more compact model (higher $\beta$, smaller
core radius) that would not reflect the true distribution of the gas,
most of which is outside the cooling flow.

Allen and Fabian \shortcite{allen98} use {\sl ASCA} and {\sl ROSAT}
HRI data to determine the temperature of A1413 in two models; one
assuming the gas is isothermal, and one also fitting for a cool
component, with the fraction of the total emission from the cool gas
normalised by the fitted mass deposition rate of the cooling
flow. These methods yield $7.5^{+0.3}_{-0.3}$ keV and
$8.5^{+1.3}_{-0.8}$~keV respectively (90\% confidence errors). The
latter value is a better estimate of the temperature of the bulk of
the cluster gas, outside the cooling flow region. Since we have
excluded the cooling flow region from our fit to the surface
brightness image, we use this higher temperature as the more
consistent model. The value of $H_0$ that would be obtained from using
the cooler temperature can be found from the scaling given in section
\ref{theanswer}. The X-ray emissivity constant was calculated assuming
an absorbing H column of $1.62 \times 10^{24}$ m$^{-2}$, a metallicity
of 0.3 solar and a temperature of 8.5~keV, giving a value of
$3.41\pm0.24 \times 10^{-69}$ $\rm counts~s^{-1}$ from 1~m$^3$ of gas
of electron density 1~m$^{-3}$ at a luminosity distance of 1~Mpc.

\begin{figure}
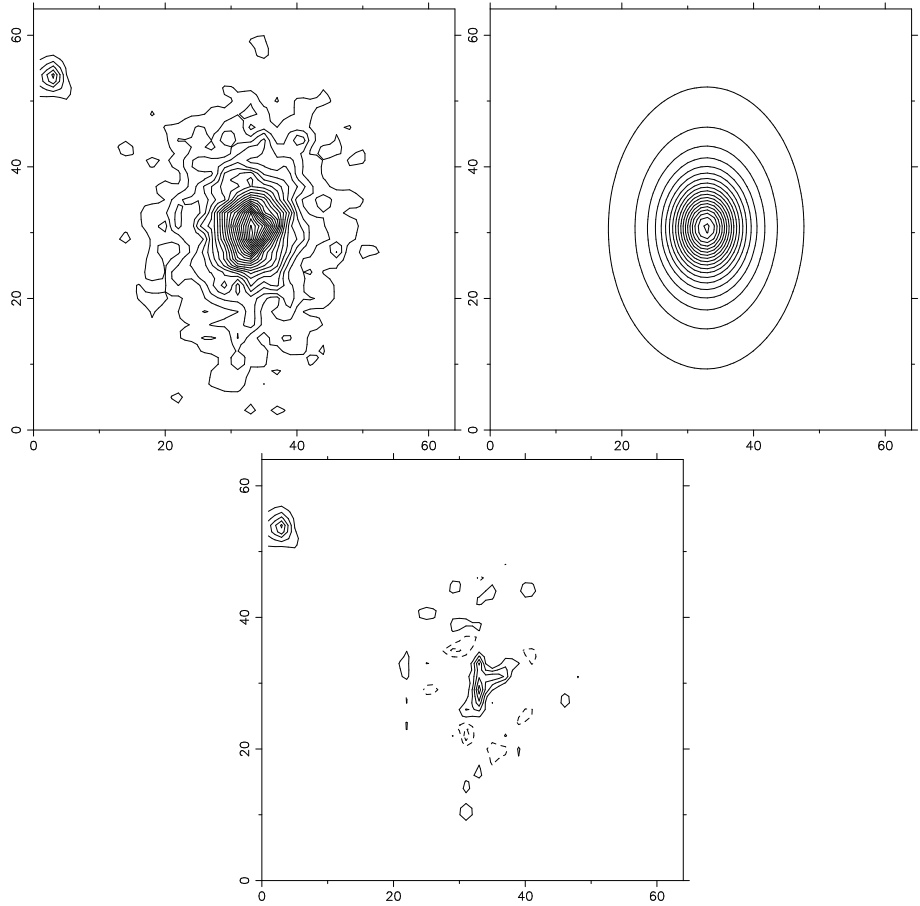

\begin{center}
\leavevmode
\epsfig{angle=270,width=60mm,file=a1413ros.eps}
\epsfig{angle=270,width=60mm,file=a1413mod.eps}
\epsfig{angle=270,width=60mm,file=a1413res8.eps}
\end{center}
\caption{\label{residuals}PSPC image, our X-ray model,
and the residuals from the fit. Contours are every 1.0 counts per
$8^{\prime\prime}$ pixel.
The cooling flow and an X-ray
point source which were not included in the fitting procedure are evident
in the residual map.}
\end{figure}

The best-fitting model parameters were $\beta = 0.57$, core radii
of~$53''$ and $35''$ with a position angle of the major axis of
$0.1^{\circ}$ and a central density $n_0 = 1.51 \times 10^4 \,
h^{1/2} \rm m^{-3}$ (assuming a core radius in the line of sight of
$43=(53\times35)^{1/2}$ arcsec), where $H_0 = 100h \, \rm
km^{-1}Mpc^{-1}$. These parameters agree well with the fit by Allen et
al \shortcite{allen95}. 
The {\sl ROSAT} image, our X-ray model and a map of the residuals
are shown in Figure~\ref{residuals}. The residual map has zero mean in the
regions included in the fitting, although the excess flux in the
cooling-flow region and a point source stand out. The Poisson noise
also rises towards the centre of the cluster as the square root of the
signal level. 
The reduced value of $\chi^2$ for this fit is 1.02; 
the mean log likelihood over the region of the fit is $L=-1.585$, and
the expected distribution of $L$, as discussed in Grainge et
al.~\shortcite{G01}, is $L= -1.585\pm0.021$. We therefore conclude that
the fit is a good one. 
There is a degeneracy between $\beta$ and the
core radii (see section~\ref{x-fitting}) 
which lies parallel to lines of constant predicted flux density observed by the
RT. Figure~\ref{degenerate} shows that acceptable fits to the X-ray
map correspond to uncertainties in the predicted SZ flux density of
$\pm 14 \, \mu$Jy and $\pm 24 \, \mu$Jy at 67\% and 95\% confidence.
Therefore we estimate that the
error in $H_0$ that 
arises solely from our X-ray fitting procedure is $2\times
{{14}\over{650}} = 4\%$.

\subsection{Relativistic Corrections}

The formula for the SZ effect quoted above (eqn~\ref{szeffect}) uses
non-relativistic approximations, which become more significant at
higher gas temperatures and higher observing frequency. Challinor and
Lasenby \shortcite{challinor98} show that a better approximation in
the Rayleigh-Jeans region (correct to second order in $T_e$) is
\[
\frac{\Delta T}{T_0} = -2y\left(1-\frac{17}{10}\frac{k_BT_e}{m_e
c^2}+\frac{123}{40}\left(\frac{k_BT_e}{m_e c^2}\right)^2\right).
\]
For A1413 this correction amounts to some 2.5\%, and we use the corrected
value in the results below.

\subsection{Fitting for \bf {$H_0$}}

\begin{figure}
\begin{center}
\psfig{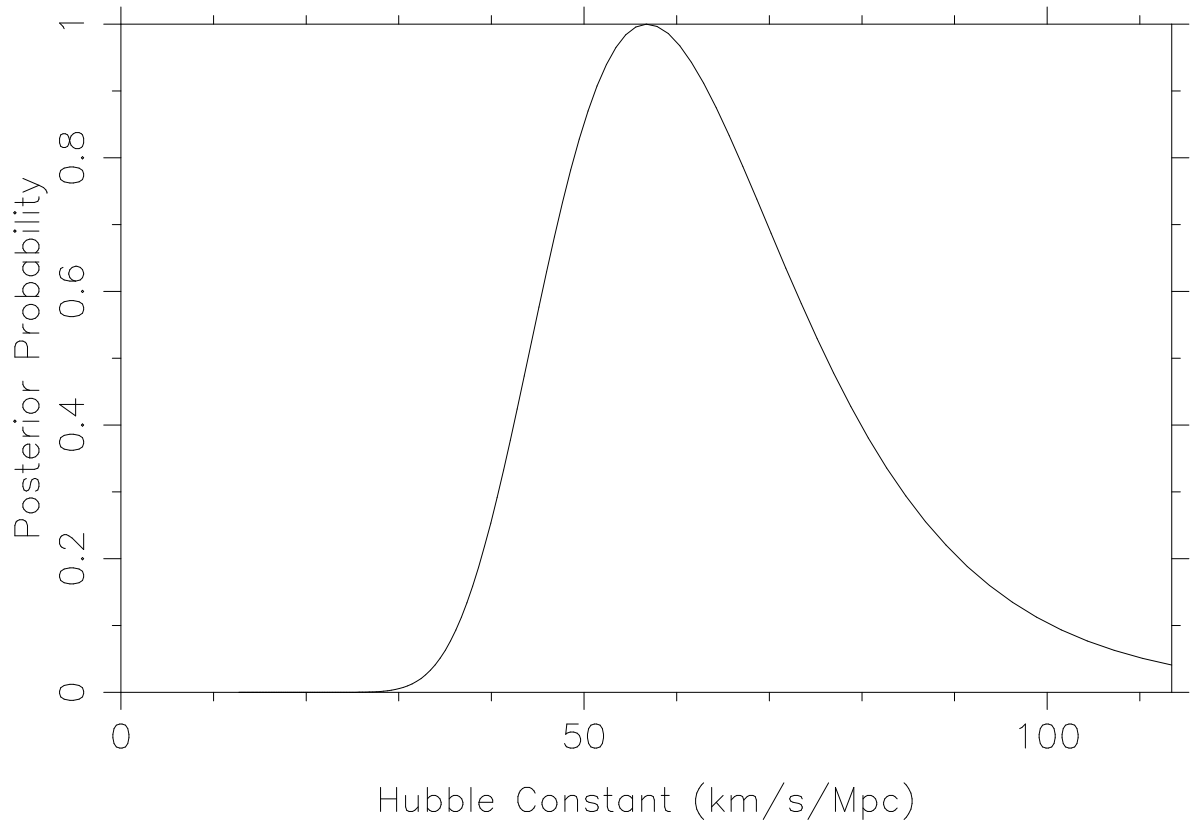}
\caption{Posterior probability plot for values of $H_0$  from the fit
of SZ decrement to the X-ray derived cluster model. The width of the
plot indicates only the errors from the visibility data.
\label{likelihood}}
\end{center}
\end{figure}

The SZ data used were those described by Grainge et al
\shortcite{grainge96}, who also describe the procedure for
subtracting the effects of radio sources within the field. We used
only the visibilities from the RT baselines shorter than 2 k$\lambda$
for the fitting, since the SZ effect is completely resolved out on
longer baselines. The mock data based on our X-ray model were 
compared with the source-subtracted data from our observations to find
best-fit values of $H_0$ and the corresponding value of $n_0$; these
were $H_0$ = $57 ^{+16}_{-13}$~$\rm km s^{-1} Mpc^{-1}$ and $n_0 = 9.0
\times 10^3$~$\rm m^{-3}$ (for an Einstein-de Sitter universe and
apply the correction for primary flux calibration discussed in
section~\ref{prim}). 
A plot of the posterior probability distribution is shown in
Figure~\ref{likelihood}. 
The central SZ temperature decrement in this model is~$825~\mu$K. The
quoted (1-$\sigma$) error on $H_0$ is just that due to the noise on
the SZ data and the fitting to the X-ray image; other sources of error
are discussed below.

\section{Other errors in the calculation of $H_0$}

There are other sources of error which may affect our estimate of
$H_0$. These arise in the radio observations, in the X-ray data, and
in the modelling process.

\subsection{Primary flux calibration of the RT}\label{prim}

The RT is primary flux-calibrated every day on either 3C~48 or 3C~286, assumed
to have flux densities in the observed Stokes' parameter {\it I}+{\it Q} of
1.70 and 3.50~Jy respectively.  These fluxes are derived from Baars et
al. \shortcite{B77}.  We estimate that the daily random error in primary
calibration is $5\%$. Since the total integration time on A1413 is 65~days,
the overall error in primary calibration due to variation during each day is
$\pm 0.6 \%$.  Since the calculated value of $H_0$ is proportional to the
square of the measured SZ flux, this leads to an error of $\pm 1.2 \%$ in
$H_0$.

Measurements by the VLA in `D' configuration have shown that the flux
densities of both 3C~48 and 3C~286 show slight variations with time
\cite{vlacalib}. This time variability is small, less than $5\%$. Measurements
in 1995, roughly contemporary with our observations, showed that the true flux
density of 3C48 was 3.4\% lower than the Baars et al value, while that of
3C286 was 1.2\% higher. Since roughly equal numbers of the individual RT
observations of A1413 were calibrated from each source, we average the
errors and take the likely systematic error as $1.1\pm2.5$\%. This results in
a reduction in $H_0$ of 2.2\%, and a further
error term of $\pm 5$\%.

\subsection{Source subtraction}\label{soursub}

We removed the effect of contaminating radio sources using higher
resolution data from both the longer baselines of the RT and from VLA
imaging of the field. However, the accuracy to which the sources can
be removed depends on the level of uncalibrated amplitude and phase
errors in the RT. From maps of secondary calibrator sources we have
found that the dynamic range of the RT is typically 100:1. Since the
total flux density of the six sources removed from the A1413 data is
3.1~mJy, the maximum spurious flux remaining after source subtraction
will be $30~\mu$Jy. However, taking into account the distribution of
the sources on the sky and the contribution each makes at the position
of the SZ decrement, we find that the maximum unsubtracted
contribution of the detected sources is only $3~\mu$Jy.

However, a more significant contribution is from the sources which
were not subtracted due to their being too faint to detect. To
calculate the confusion noise contribution from unsubtracted sources,
we take the $8.4~$GHz $\mu$Jy source counts of Windhorst et al
\shortcite{windhorst}, and extrapolate to 15~GHz using an effective
spectral index calculated from their 5--8.4~GHz spectral
measurements. Integrating the source counts over the beam area
corresponding to our shortest baseline, from zero flux density up to
our source detection limit of $120~\mu$Jy, and taking into account the
increased source population in clusters compared to the field
\cite{cooray}, we find a residual confusion noise of $60~\mu$Jy, which
adds in quadrature with the thermal noise. However, since we estimate
our thermal noise from the scatter in the visibilities, the confusion
noise is already included in our noise estimate, and does not need to
be added in again.

\subsection{X-ray emission constant}

The X-ray emission constant, $K$, used in simulating the X-ray data is
dependent upon the telescope detector response in the appropriate
frequency band, the K-correction for redshift, the cluster temperature
and the absorption due to Galactic hydrogen. We estimate that this
constant could be in error by $\pm 7\%$ leading to a $\pm 7\%$ error
in $H_0$.

\subsection{Extended radio emission}

Some clusters of galaxies are known to have diffuse (`halo') emission
on arcminute scales. Any such emission at 15~GHz would be resolved out
on the longer RT baselines used for source subtraction, and would
therefore contaminate the SZ data. Observations at 1.4~GHz from the
NVSS and FIRST surveys show that there is 1.9~mJy of flux associated
with the central cluster galaxy which is extended on scales of
$\approx 45''$, but none on larger scales. This source was detected
(and indeed slightly resolved) in the RT longer baselines and was
removed from the data (source 3 in Grainge et al
\shortcite{grainge96}). The spectral index of this source
$\alpha^{15}_{1.4}$ is 0.8, typical for a cluster galaxy but quite
atypical for a diffuse halo source, where one would expect
$\alpha^{15}_{1.4} > 1.4$ \cite{H82}. Also, halo emission is unknown
in clusters with strong cooling flows such as A1413. We therefore
conclude that there is no evidence for any unsubtracted diffuse
emission in A1413.

\subsection{X-ray estimate of gas temperature}

The temperature we use has a quoted 90\% confidence error of
$^{+15}_{-9}\%$. Converting this to a 1-$\sigma$ errors for
consistency gives $^{+10}_{-6}\%$, which will lead to a
$^{+21}_{-12}\%$ error in $H_0$.

\subsection{Cluster ellipticity} 

X-ray images of clusters show that clusters are typically elliptical with
ellipticities of up to 1.2:1 being common.  Therefore one cluster is not
sufficient to provide a definitive value of $H_0$.  Making the necessary
corrections to equation \ref{calcH0}, one finds that the calculated value $
H_0^{\rm calc}$ is related to the true value by:
\[  {H_0^{\rm calc}}\propto {H_0}^{\rm true}  {{l_{\perp}}\over{l_{\parallel}}},\]
for a cluster with line-of-sight depth $l_{\perp}$ and observed
diameter $l_{\parallel}$.  Therefore the derived $H_0$ from a sample
where there is no bias in the orientation of clusters will form a
distribution with a geometric mean close to the true value. It is thus
necessary to observe such a sample of clusters in order to reduce the
uncertainty due to this effect.  An X-ray surface-brightness limited
sample will be biased towards selecting clusters elongated towards
us. However, an X-ray sample selected on X-ray flux will be
independent of cluster orientation.

In a case such as the present where one is estimating $H_0$ from a {\em
single} cluster, it is appropriate to make some allowance for
orientation uncertainty. For A1413 we have assumed that the
line-of-sight depth through the cluster is the mean of the two
elliptical axes in the plane of the sky. This is an unbiased
estimator of the true depth, assuming the cluster is drawn from an
unbiased population~\cite{grainger2001}. (Sulkanen
\shortcite{sulkanen} finds a similar result taking the arithmetic
mean.) 
In order to estimate the error introduced by this assumption we
compare model distributions of elliptical clusters with
those observed in a sample of ROSAT-selected clusters over a redshift
range of $z=0.1$--0.5. We find that the effect of cluster
ellipticity adds an error of $14\%$
for each cluster to the calculated value of $H_0$. 

\subsection{Effect of temperature structure}\label{coldhalos}

Considerations of hydrostatic equilibrium, as well as observations on nearby
clusters \cite{markevitch} show that the isothermal assumption breaks down at
radii much bigger than $r_c$, the temperature falling by a factor of about 2
at $6r_c$. This can in principle have a significant effect on the derived value of $H_0$, tending to bias it downwards \cite{inagaki,roettiger}. For A1413, one core diameter corresponds to 2 arcminutes, which is
approximately the scale to which the RT is sensitive. Gas on larger scales
than this does not affect the flux density seen by the RT, despite the effect
it has on the SZ central decrement. {\sl ASCA} observations do encompass larger
scales, but the temperature measurement is necessarily emission-weighted, and
hence is a good measurement of the temperature of the gas that the RT sees.
We therefore conclude that SZ observations by the RT are not going to be
significantly affected by the presence of a cold halo of gas, and so we can
calculate $H_0$ without requiring knowledge of the gas temperature at large
radii. A future paper will deal with this effect in more detail.

\subsection{Clumping of the intracluster gas}

X-ray imaging on nearby, large angular size clusters has found no
evidence for clumping, and has placed constraints on the degree of
clumping which may be present~\cite{FCEM94}. However it has been
suggested that the scatter in the temperature--luminosity correlation
for clusters is due to the existence of clumping below this level and
further that the degree of clumping is correlated with the strength of
the cooling flow \cite{KTK91}. If the clumps are close to being in
pressure equilibrium with their surroundings the clumping will have no
effect on the SZ signal.  The X-ray luminosity will increase but the
effect of this in the $H_0$ estimate will largely be cancelled due to
the decreased emission-weighted temperature.  The net result would be
a modest underestimate of $H_0$. Initial simulations indicate that
fractionation leading to an underestimate of $2\%$ in $H_0$ would be
consistent with the temperatures calculated by Allen and
Fabian~\shortcite{allen98}.

\subsection{The kinetic SZ effect}

Bulk motion of the cluster gas along the line of sight will result in a
further distortion of the CMB spectrum due to the Doppler effect. 
Watkins \shortcite{w97} finds that the 1-d rms peculiar velocity of
clusters is $\sigma_{v_z} = 265^{+106}_{-75}~\rm{km~s^{-1}}$ (at
$90\%$ confidence). At 15~GHz, a cluster with a peculiar velocity of
265~km~$\rm s^{-1}$ will have a kinetic SZ effect with a magnitude
$2.5\%$ of that due to the thermal SZ effect and so introduce a $5\%$
error in the determination of $H_0$ from a single cluster.  Since
randomly selected clusters will have random directions of peculiar
motion, the error introduced into $H_0$ calculations can be reduced by
averaging over a large sample of clusters.

\subsection{Radio bremsstrahlung}

It is possible that the electron cluster gas could emit sufficient
thermal bremsstrahlung radiation at 15~GHz to significantly affect the
SZ decrement which we observe. We calculated the emissivity of the hot
thermal component of the gas using our cluster model, taking the Gaunt
factor to be 10 at a frequency of 15~GHz. We find that the total
integrated flux density from the entire cluster is $15~\mu$Jy. This
emission will be distributed like the X-ray emission, and the
resolving-out of our shortest baseline means that we will observe only
about 20\% of it. Since the emissivity due to brem\-sstrah\-lung is
proportional to $n^2$, it is possible that the effect of thermal
bremsstrahlung will be much greater in a cluster which has a strong
cooling flow and thus a high gas density at the cluster centre
\cite{tarter,schlickaiser}. We investigated this possibility by using
a good empirical fit to the form of a cooling flow, determined from
X-ray observations, allowing the density to increase as $1/r$ for $r$
within the cooling flow radius of 100--200~kpc until $r = 10$~kpc. We
find that this cooling flow contributes a further $4~\mu$Jy of flux
density.  Therefore we conclude that even for this worst case example,
and assuming that none of the cooling-flow flux is resolved out by the
RT, the effect of thermal bremsstrahlung will be to overestimate $H_0$
by up to $2\%$.

\subsection{Effects of gravitational lensing}

The strong gravitational potential of the cluster will inevitably lead to
lensing of background radio sources in the field.  Refregier and Loeb
\shortcite{RL97} have calculated the systematic bias this introduces into
$H_0$ calculations.  Assuming $\beta = 0.6$, using a $2'$ beam and subtracting
sources above $100~\mu$Jy, we find that approximately $0.5\%$ of the SZ
temperature decrement may be due to this effect. This leads to an
underestimate in our value to $H_0$ of $1\%$.

\subsection{Combining the errors}
\label{theanswer}

All the errors listed above are summarised in table \ref{table}. Writing the
result as a function of the systematic quantities concerned, we find
\[
H_0 = 57^{+16}_{-13} \times \left( {{{1.64~\rm{Jy}}\over{S_{3C48}}} +
{{3.54~\rm{Jy}}\over{S_{3C286}}}\over{2}} \right)^2 \times
\left({{3.41\times10^{-69}}\over{K}} \right) \cr \times \left
( {{k_BT_e}\over{8.5~\rm{keV}}} \right)^2 \times
\left({{\theta_z}\over{57''}}\right) \times  \left(1+0.085{{v_{\rm
z}}\over{1000~\rm{km~s^{-1}}}} \right)^2 \cr \times
\left(1-{{S_{\rm{brems}}}\over{640~\mu\rm{Jy}}} \right)^2 \times
\left(1+{{S_{\rm{lens}}}\over{640~\mu\rm{Jy}}} \right)^2 \rm \, km \, s^{-1}
\, Mpc^{-1}
\]
where $S_{3C48}$ and $S_{3C286}$ are the true primary calibrator flux
densities (I+Q), $K$ is the X-ray emissivity constant as defined in
section \ref{a1413xfit}, $T_e$ is the gas temperature, $\theta_z$ the
line-of-sight core radius, $v_z$ the line-of-sight peculiar velocity,
and $S_{\rm brems}$ and $S_{\rm lens}$ are the flux densities due to
bremsstrahlung emission and lensed background sources
respectively. Inserting our best estimates for the errors in these
quantities, we find in the worst case, where the errors conspire to
add together linearly, $H_0 = 57^{+58}_{-27}$~$\rm km s^{-1}
Mpc^{-1}$. However, since these errors are all independent of each
other, we can justifiably combine them in quadrature with the random
error. Doing this we obtain $H_0$ = $57^{+23}_{-16}$~$\rm km s^{-1}
Mpc^{-1}$. The largest errors are due to the uncertainty in X-ray
temperature, the noise on the SZ measurement and uncertainty in
cluster line-of-sight depth.

\section{Conclusions}

We have described our methods of determining the Hubble constant from
SZ and X-ray observations of clusters of galaxies, giving careful
consideration to sources of both random and systematic error. We have
used observations of the cluster A1413 to estimate $H_0$. We conclude
that:

\begin{enumerate}

\item The error due to unsubtracted radiosources is comparable to the
thermal noise, and adds in quadrature with it; our estimate of the
noise from the scatter in the visibilities includes both effects.

\item Fitting the X-ray emission using a King model gives
a degeneracy in the $\beta, \ \theta$ plane. The predicted flux
density that the RT will observe based on any of these good-fit models
is almost constant, with the result that the error in $H_0$ that is
introduced solely by our X-ray fitting is small, approximately~$\pm 4\%$. 

\item There is a random uncertainty in the calculation of the X-ray
emissivity of the cluster gas which we estimate to be 7\%.

\item The orientation of any non-spherical cluster causes a random
uncertainty in the value of $H_0$ of some 14\%. This error may be
reduced by observing a complete sample of randomly-oriented clusters. 

\item A colder cluster atmosphere at large radius will not affect our
determination of $H_0$ much, but clumps of cool gas within the cluster may
result in an underestimate the value of $H_0$ of up to 2\%.

\item The kinetic SZ effect will cause a random uncertainty of
approximately 2.5\% the value of the thermal SZ effect. This error may
be reduced by observing a complete sample of randomly-oriented
clusters.

\item We estimate that the contribution of thermal bremsstrahlung at
radio wavelengths will cause an overestimate of $H_0$ of 2\% in the
worst case.

\item Gravitational lensing will cause the {\em over}-subtraction of
flux from contaminating background radiosources; this is a small
effect causing an underestimate in  $H_0$ of about 1\%.

\item The dominant sources of error are thus the thermal/confusion
noise in the SZ measurement, the temperature of the gas, and the
unknown line-of -sight depth. Combining these sources of random and
systematic error in quadrature, we determine from our observations of
A1413 a value of $H_0 = 57^{+23} _{-16}$~$\rm km s^{-1} Mpc^{-1}$.

\end{enumerate}

\section{Acknowledgements}

We thank the staff of the Cavendish Astrophysics group who ensure the
continued operation of the Ryle Telescope. Operation of the RT is
funded by PPARC. AE acknowledges support from the Royal Society; WFG
acknowledges support from a PPARC studentship.

\clearpage

\begin{table}
\begin{center}
\begin{tabular}{l|l}
Source of error & \% error in $H_0$\\
\hline
Noise on SZ & $ \pm 28 \%$\\
Fitting to X-ray map & $ \pm 4 \%$\\
RT primary calibration & $\pm 5\% $\\
Source confusion & $\pm 16\% $\\
X-ray emission constant &$ \pm 7 \% $\\
Cluster gas temperature & $+20/-12\%$ \\
Ellipiticity &$ \pm 14 \%$ \\
Clumping &  $+2 \%$ \\
Kinetic SZ & $\pm 5 \%$ \\
Radio bremsstrahlung &$ -2 \%$ \\
Gravitational lensing of background sources & $+1 \%$ \\
\end{tabular}
\end{center}
\caption{Error budget for $H_0$ determination from A1413. For each
source of error a $+$ sign
indicates that the true value of $H_0$ may be more than the calculated
value.}
\label{table}
\end{table}

\clearpage

\clearpage

\clearpage

\clearpage

\clearpage

\clearpage

\clearpage

\end{document}